\documentclass[preprint,showpacs,preprintnumbers,amsmath,amssymb,superscriptaddress]{revtex4-1}
\usepackage{graphicx}
\usepackage{siunitx}
\usepackage{soul}
\usepackage{rccol}
\newcolumntype{d}{R[.][.]{1}{2}}
\newcolumntype{e}{R[.][.]{1}{1}}
\usepackage{natbib}
\usepackage[normalem]{ulem}
\def\fixmacro#1{#1}

\def\employ{{
employ}}

\begin{document}
\title{Comparison of the periodic slab approach with the finite cluster ansatz
  for metal-organic interfaces at the example of PTCDA on Ag(110)}

\author{Jaita Banerjee} 
\altaffiliation[Present address: ]{University and Institute of Advanced
  Research, Koba Institutional Area, Gandhinagar - 382007, Gujarat, India} 

\author{Stefan Behnle}
\affiliation{Institut f\"{u}r Physikalische und Theoretische Chemie, Auf der
  Morgenstelle 18, Universit\"{a}t T\"{u}bingen, 72076 T\"{u}bingen, Germany}

\author{Martin C.\ E.\ Galbraith} 
\altaffiliation[Present address: ]{Max-Born Institute, Berlin Germany.}
\affiliation{Fachbereich Chemie,
  Philipps-Universit\"{a}t Marburg, Hans-Meerwein-Str. 4, D-35032 Marburg,
  Germany} 

\author{Hans-Georg  Mack} 
\affiliation{Institut f\"{u}r Physikalische und Theoretische Chemie,
  Auf der Morgenstelle 18, Universit\"{a}t T\"{u}bingen, 72076 T\"{u}bingen,
  Germany} 

\author{Volker Settels} 
\altaffiliation[Present address: ]{BASF SE
  GNC/M - B001 Ludwigshafen Germany} 
\affiliation{Institut f\"{u}r Physikalische
  und Theoretische Chemie, Universit\"at W\"urzbung, Emil-Fischer-Str.\ 42,
  97074 W\"{u}rzburg, Germany. } 

\author{Bernd Engels}
\affiliation{Institut f\"{u}r Physikalische und Theoretische Chemie,
  Universit\"at W\"urzbung, Emil-Fischer-Str.\ 42, 97074 W\"{u}rzburg,
  Germany. } 

\author{Ralf Tonner} 
\email{tonner@chemie.uni-marburg.de}
\affiliation{Fachbereich Chemie,
  Philipps-Universit\"{a}t Marburg, Hans-Meerwein-Str. 4, D-35032 Marburg,
  Germany} 

\author{Reinhold F. Fink} 
\email{reinhold.fink@uni-tuebingen.de}
\affiliation{Institut f\"{u}r Physikalische und
  Theoretische Chemie, Auf der Morgenstelle 18, Universit\"{a}t T\"{u}bingen,
  72076 T\"{u}bingen, Germany}

\begin{abstract}
  We present a comparative study of metal-organic interface properties
  obtained from dispersion corrected density functional theory calculations
  based on two different approaches: the periodic slab supercell technique and
  cluster models with 18 to 290 Ag atoms. Fermi smearing and fixing of cluster
  borders are required to make the cluster calculation feasible and realistic.
  The considered adsorption structure and energy of a PTCDA molecule on the
  Ag(110) surface is not well reproduced with clusters containing only two
  metallic layers. However, clusters with four layers of silver atoms and
  sufficient lateral extension reproduce the adsorbate structure within 0.02
  \AA\ and adsorption energies within 10\% of the slab result. A consideration
  of the computational effort shows that the cluster approach is a competitive
  alternative to methods using periodic boundary conditions and of particular
  interest for research at surface defects and other systems that do not show
  periodic symmetry.
\end{abstract}

\date{\today}
\maketitle

\section{Introduction}

Interfaces formed when large organic $\pi$-conjugate molecules adsorb on
metallic substrates represent crucial functional parts in a variety of nano
and micro, electronic and optoelectronic devices, such as organic solar cells,
organic light emitting diodes, organic field-effect transistors  and other
devices \cite{Scott1999115, Forrest2004}. The importance of these interfaces
is that they are often essential for the generation, injection and
transportation of charges in the device \cite{scott:521, Santato2010,
  Duhm2008111}. The bonding nature of the molecule to the surface is of
particular interest as it affects these properties substantially
\cite{Tseng2010, ADMA:ADMA200802893}.

Several interesting studies of organic-metal interfaces are found in the
literature, where the interactions of a variety of organic molecules
(pentacene, perylene derivatives, phthalocyanines) have been studied on the
low index planes of noble metals, namely Ag, Cu and Au \cite{4866097,
  PhysRevB.81.045418, PhysRevB.75.045401, Tautz2007479, PhysRevB.86.235431,
  PhysRevB.68.115428,PhysRevB.87.165443, PhysRevB.71.205425,
  PhysRevB.84.125446}. In this paper we will restrict the discussion to the
adsorption of Perylene-3,4,9,10-tetracarboxylic dianhydride (PTCDA) on the
Ag(110) surface which is a prototype system as its adsorption characteristics
have been very well studied \cite{Temirov2006, Tautz2007479,
  PhysRevB.73.165408, PhysRevB.61.16933, PhysRevB.86.045417,
  PhysRevLett.104.233004, PhysRevB.86.235431, Scholz_2009_long_paper,
  Zou20061240,Mer13a,Mau16}. Besides detailed information on the lateral
orientation of the molecule upon the metal surface, x-ray standing wave (XSW)
measurements provide accurate information about the distance between the
individual atoms of the molecule from the surface
\cite{PhysRevB.86.045417,Mer13a}.

Density Functional Theory (DFT) \cite{Hohenberg_Kohn_Theorem,
  PhysRev.137.A1697} is one of the most widely used tools for studying metals
and metal surfaces due to its ability for providing a realistic description of
the electronic structure of these systems. Unfortunately, local
\cite{gross1985local, casida1998molecular, Wes02} or semi-local
\cite{Perdew_Wang_1992, PBE_1996} density functionals are not able to properly
reproduce London dispersion interactions which are significantly affecting the
structure of aggregates in general and thus also of metal adsorbate systems.
To address this issue, several techniques have been designed 
\cite{PhysRevB.28.1809, PhysRevLett.102.073005, PhysRevLett.108.146103,
  PhysRevB.28.1809, Grimme_D_2004, Grimme_D2_2006, Grimme_D3_2010,
  Grimme_2011_Review, Grimme_2011_damping_function,Mau16}. It was shown that
the inclusion of dispersion corrections is crucial in the treatment of large
$\pi$-systems on noble metal substrates
\cite{Cap07,Tkatchenko_2010,Scholz_short_paper,Scholz_2009_long_paper,
  PhysRevLett.99.176401,Mau16}. Of these, the DFT-D3(BJ) empirical scheme
developed by Grimme {\textit{et al.}}~\cite{Grimme_2011_damping_function}
performs remarkably well in predicting the structure and adsorption energy of
organic molecules on surfaces~\cite{Mer13a,PhysRevB.86.235431,Ton16}.

The most common theoretical approach to metal-organic interfaces is the
periodic slab-supercell approach (see e.g.
\cite{Puschnig_2011,PhysRevLett.99.176401,Ton16,Yim13}) where the adsorbate is
placed on the surface and periodic boundary conditions are applied. The
resulting system replicates a physical surface-adsorbate system by an infinite
stack of equally spaced metallic slabs. The slabs are rather thin periodic
metallic surfaces with molecules adsorbed in a two dimensional periodic
arrangement. Tuning the distance between two adjacent periodic images allows
to control the interaction between neighbouring slabs while 
modulating the size of the surface unit cell
makes it possible to control the interaction
between neighbouring adsorbate molecules.

As an alternative to the slab model, the cluster approach can be {\employ}ed by
cutting several metal atoms out of the surface and placing a molecule upon it.
This means that the metal surface is represented by a brick of atoms in the
vacuum \cite{volker}. This model has also been used frequently, e.g.\ for
adsorbates on silver~\cite{Scholz_short_paper,Scholz_2009_long_paper,Cap07}
and other metals~\cite{Jac06,Cap07,Bag10,Ket12,Sch13c,Nig14}. Generally it is
significantly more difficult to design a proper cluster than to set up a
periodic slab calculation as it is not clear how the lateral extension of the
cluster and the corresponding borders influence the metal adsorbate
properties. Another challenge for the cluster approach is to find appropriate
orbital occupations. Due to small energy gaps it is generally unclear which
orbitals should be occupied and which spin multiplicity is appropriate to
describe the electronic structure of the metal surface \cite{Ket12,Jac06}.

However, cluster calculations have some advantages: They are usually conducted
with quantum chemical program packages using local Gaussian basis sets whereas
continuous plane
wave basis sets are most commonly {\employ}ed for the slab model. For a given
system the number of local basis functions required to expand the electronic
wavefunctions is generally significantly smaller than the corresponding number
of plane wave functions. The latter  allow for very efficient
evaluation of the Coulombic electron repulsion energy \cite{Haf08}
which overcompensated
the disadvantage of dealing with a higher number of basis functions for a long
time. However, efficient implementations of DFT in Quantum Chemical program
packages are available \cite{Eichkorn_1997,Sierka_2003}. These  make use of the
local nature of the basis functions and thus may turn the
situation. Furthermore, the cluster approach is applicable to charged
adsorbates \cite{Bag10} as well as other  non-periodic
systems and -- in principle -- it allows to apply accurate wave function
based methods. 

The motive of this work is to investigate whether structural and energetic
properties of an adsorbate upon a metal surface converge to the results of a
corresponding slab model by systematically increasing the cluster size. This
comparison requires that the atomic and electronic structure of the metal is
set up in the same way. The latter is achieved in our approach by {\employ}ing
the same structure at the cluster border and by using a pseudo Fermi
\cite{Rab99} type
occupation of the discrete orbitals. Basis set convergence is investigated as
well as the performance of different dispersion correction schemes.

The paper is organised as follows; in section II the computational details of
the two approaches are discussed. In section III, we present the results
obtained from the slab-supercell calculations and the cluster calculations,
and a comparison between them. In section IV we summarise the work and discuss
how the cluster ansatz may be improved and {\employ}ed in future work.

\section{Methods}
\label{sec:methods}

\subsection{Structural Information}
\label{sec:gener-struct-setup}

All calculations {\employ}ed DFT using  the generalised gradient
approximation (GGA) Perdew-Burke-Ernzerhof (PBE) functional \cite{PBE_1996}.
Three types of dispersion schemes were used: (i) without applying any
dispersion correction (noD), (ii) applying the D2~\cite{Grimme_D2_2006}
correction and (iii) the
D3(BJ)~\cite{Grimme_2011_Review,Grimme_2011_damping_function} correction which
uses the rational damping of the dispersion contribution to finite values for
small inter atomic distances as proposed by Becke and Johnson \cite{Bec07}.

As described below, the slab-supercell approach requires only information of
the bulk structure while our cluster approach needs also the relaxation
pattern of the metal atoms near to the surface. This structural information
was obtained with the Vienna Ab Initio Simulation Package (VASP) code which
was also {\employ}ed for the slab-supercell calculations. All VASP
calculations were performed using 
plane wave basis sets and the projector augmented wave
technique for treating core electrons efficiently~\cite{PhysRevB.47.558,
  PhysRevB.49.14251, PhysRevB.54.11169, Kresse199615}. The energy cut-off was
set to 340 eV as this was found to be sufficient to converge the bulk lattice
constant within $<1$~\%. For further details see supporting information of
Ref.~\cite{Ton16}. The Methfessel-Paxton second order smearing technique
\cite{PhysRevB.40.3616} with a width of \SI{0.2}{eV} was used.
For the determination of the bulk structure, the
Brillouin zone was sampled with a Monkhorst-Pack \cite{PhysRevB.13.5188}
distribution of $11\times 11\times 11$ {\textbf{k}}-point mesh.

\subsection{Slab-supercell Approach}
\label{sec:slab-superc-appr}

All slab-supercell calculations were performed  with a rectangular 
\begin{tiny}
  $\scriptstyle\left(
    \begin{array}{rr}
      3&2\\
      -3&2
    \end{array}
  \right)$
\end{tiny}
surface unit cell. The slab was chosen to have four atomic layers, the two
lowermost layers were kept fixed at the bulk equilibrium structure obtained
with the respective dispersion correction while the remaining two were
relaxed. A sufficient vacuum separation of 20 \AA\ was used for all
calculations and the Brillouin zone was sampled with a $3\times 3\times 1$
{\bf k}-mesh. The nearest neighbour Ag-Ag distances resulting from the
structural optimisation of the plain slabs are collected in
Tab.~\ref{tab:silver-distances}.

The adsorbate system was obtained by placing a PTCDA molecule on the Ag(110)
surface in the experimentally determined lateral position
\cite{Gloeckler_1998_1,Boehringer_1998,Mer13a} where the acyl and anhydride
oxygen atoms are located silver atoms (atop sites) while the perylene core is
located between two Ag rows.

The adsorption energy of the molecule to the Ag(110) surface was obtained by
\begin{equation}
E_\text{ads} = E_{PTCDA/Ag(110)} - (E_{Ag(110)} + E_{PTCDA}) ,
\label{eqn-ads}
\end{equation} 
where $E_{PTCDA/Ag(110)}$ is the total energy of the system when a PTCDA
molecule is placed on the Ag(110) surface. $E_{Ag(110)}$ is the total energy
of the bare Ag(110) surface and $E_{PTCDA}$ is the total energy of a PTCDA
molecule in a large box without inter-molecular interactions.

\subsection{Finite Cluster Approach.}
\label{sec:clust-dft-appr}

\subsubsection{Constructing Ag(110) Clusters.}
\label{sec:constr-ag110-clust}

The Ag(110) ``surface'' clusters were constructed from the optimised bare
Ag(110) slab-supercell structures at the PBE-noD, -D2 and -D3(BJ) level (see
Tab.~\ref{tab:silver-distances} for the respective structural parameters).
They contain two or four layers of Ag atoms
 along the (110) direction. A layer consisting of $n$ atoms in
the $(1\bar 1 0)$ and $m$ atoms in the (001) direction is designated as
$(n\times m)$. All clusters are constructed such that the
uppermost (first) layer of $(n\times m)$ Ag atoms is followed by a larger
second $((n+1)\times (m+1))$ layer. Thus, the Ag-atoms contacting the adsorbate
are relatively well embedded by the second layer atoms. This construction
pattern avoids ``naked'' atoms at the surface and provides rather stable
cluster surfaces. The third and fourth layers (if present) contain $(n\times
m)$ and $((n-1)\times (m-1))$ Ag atoms.

\begin{figure*}
\includegraphics[width=\textwidth]{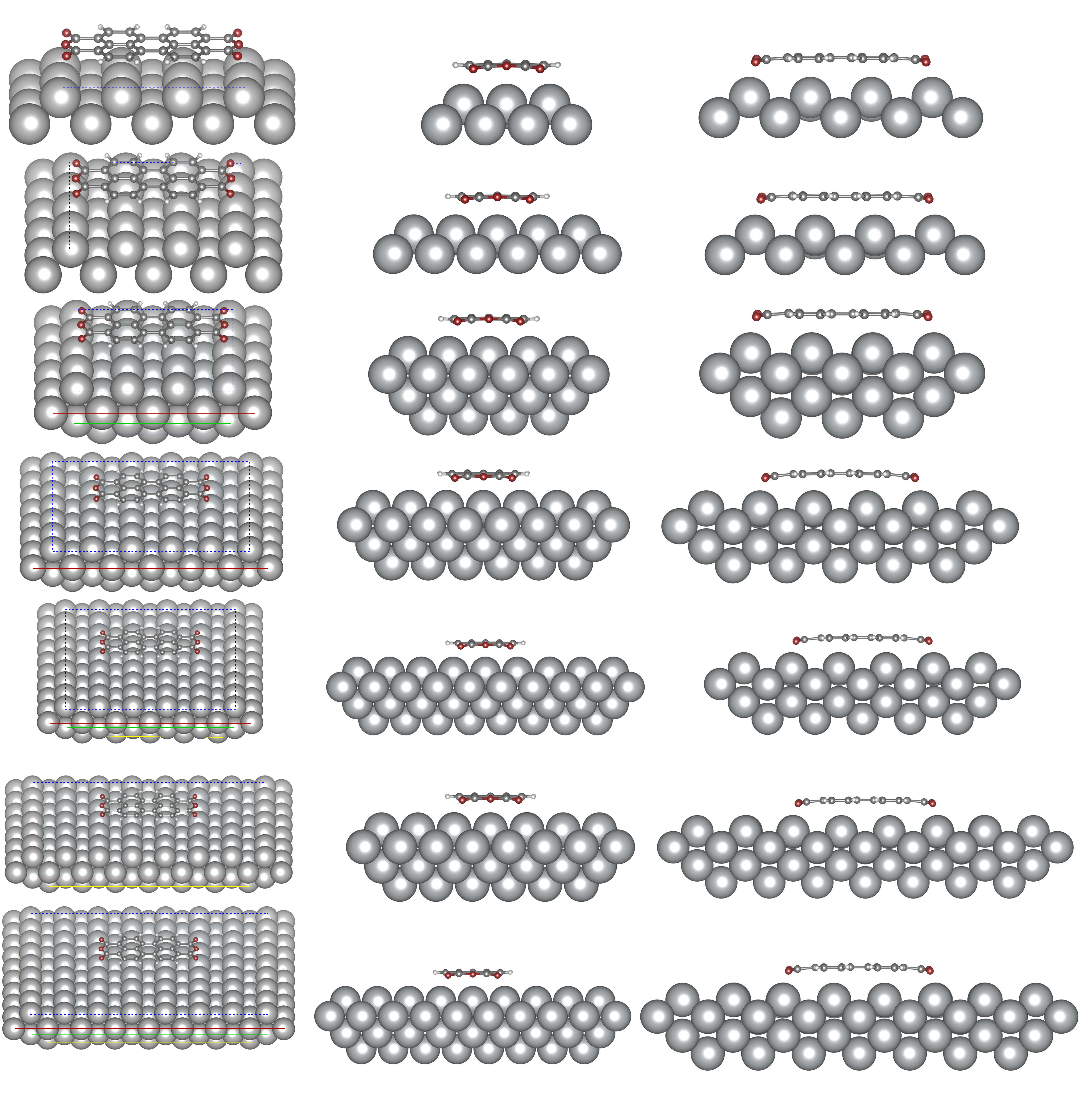}
\caption{Top to bottom: PTCDA on Ag(110) clusters consisting of 32, 50, 82,
  170, 218, 226 and 290 atoms. Column 1: Initial configuration of PTCDA on the
  Ag(110) clusters. The blue rectangles border the topmost surface layer, the
  red, green and yellow lines is a guide to denote the 2nd, 3rd and 4th layers
  in all clusters except the 2 layered 32 and 50 atom clusters. Column 2 and
  Column 3 shows the side view along the molecular short axis and long axis
  respectively, of exemplary optimized structures. The molecule shows
  significant 
  bending along the long axis for the larger clusters, starting from the 170
  atom Ag(110) cluster. 
}
\label{figure1}
\end{figure*}
The smallest cluster, Ag32, has two atomic layers with a 
$(3\times4)$ top layer, and consists of 32 atoms. The second
cluster of 50 atoms consists of a $(5\times4)$ top layer and has also two
layers. The other clusters all contain four layers with the following top
layers: 
Ag82  $(5\times 4)$, 
Ag170 $(7\times 6)$,
Ag218 $(9\times 6)$,
Ag226 $(7\times 8)$,
Ag290 $(9\times 8)$.




For the clusters with two atomic layers, the lowermost layer was kept fixed,
while the two lowermost layers were fixed for the four layer clusters just as
in the periodic slab calculations. We adopt two schemes for fixing the border
atoms of the first and the second top layers, which we refer to as
{\fixmacro{fix12}} - when the border atoms of both the top and the second
layers are kept fixed and {\fixmacro{fix2}} - when only the border atoms of
the second layer are kept fixed. 


\subsubsection{Calculation details.}

All cluster calculations were performed with the program package
TURBOMOLE~\cite{TURBOMOLE,Ahlrichs_1989_TURBOMOLE,v_Arnim_TM_parallel}
{\employ}ing the
resolution of the identity (RI) method
\cite{Eichkorn_1995,Eichkorn_1997,Vah93} and the multipole accelerated RI-J
(MARI-J) approximation \cite{Sierka_2003} which are known to speed up
calculations without introducing significant errors. We used the
def-SV(P)~\cite{Sch92a}, def2-TZVP~\cite{Wei05} and the def2-QZVP~\cite{Wei05}
basis sets with the corresponding auxiliary (RI) basis
sets~\cite{Eichkorn_1997,Wei06} and ecp-28 effective core potentials for the
silver atoms \cite{Andrae_1990_ECP}. We {\employ}ed the m3 grid as implemented
in TURBOMOLE \cite{Eichkorn_1997} to evaluate the energy expression and
exploited the $C_{2v}$ symmetry of the systems. A rather large damping
parameter~\cite{ESQCI99} in the range between 3.7 and 1.05 had to be used in
order to guarantee convergence of the self consistent field (SCF) iterations.
A total energy convergence criterion of \SI{1e-5}{E_{h}}, a
Cartesian gradient norm criterion of \SI{1E-4}{au} and an SCF convergence
criterion of \SI{1e-6} (see Ref.~\cite{ESQCI99} for details on these
parameters) was sufficient to converge structural parameters to
\SI{0.01}{\angstrom} and adsorption energies to \SI{0.01}{eV}. Further details
are given in the supporting information of this article.
The cluster computations were carried out using standard workstations with 4
cores or cluster nodes with 8 cores and up to 64 GB RAM.

In a quantum chemical program such as TURBOMOLE it is common to occupy
orbitals with integer numbers of electrons. This turned out to be rather
impractical for the cluster calculations as the HOMO-LUMO gap becomes very
small such that orbitals near this gap frequently change their occupation
within the SCF or structure convergence iterations. We found a straightforward
albeit computationally demanding solution by performing a pseudo Fermi
\cite{Rab99} smearing of the occupation of the cluster orbitals where a
constant electronic temperature of \SI{300}{K} was chosen. This was
incorporated in context with spin restricted Kohn-Sham determinants. For the
larger clusters, several orbitals had fractional occupation numbers in the
range between 0.1 and 1.9 electrons even in the converged orbitals. As Fermi
smearing spoils the convergence acceleration method implemented in TURBOMOLE,
in many cases several hundred SCF iterations were required to converge the
electronic structure of a given cluster model. Nevertheless, the efficiency of
the RI and MARI-J approximations allowed to conduct structural optimisations
of even the largest clusters on our relatively simple computing resources within
a few days. The wall clock time for the full structural optimisation of the
PTCDA@Ag290 system with PBE-D3(BJ) and the def2-TZVP basis set was in the
order of 20 days.

\subsection{Definition of Distances.}

\begin{figure}
\centering
\includegraphics[width=\columnwidth]{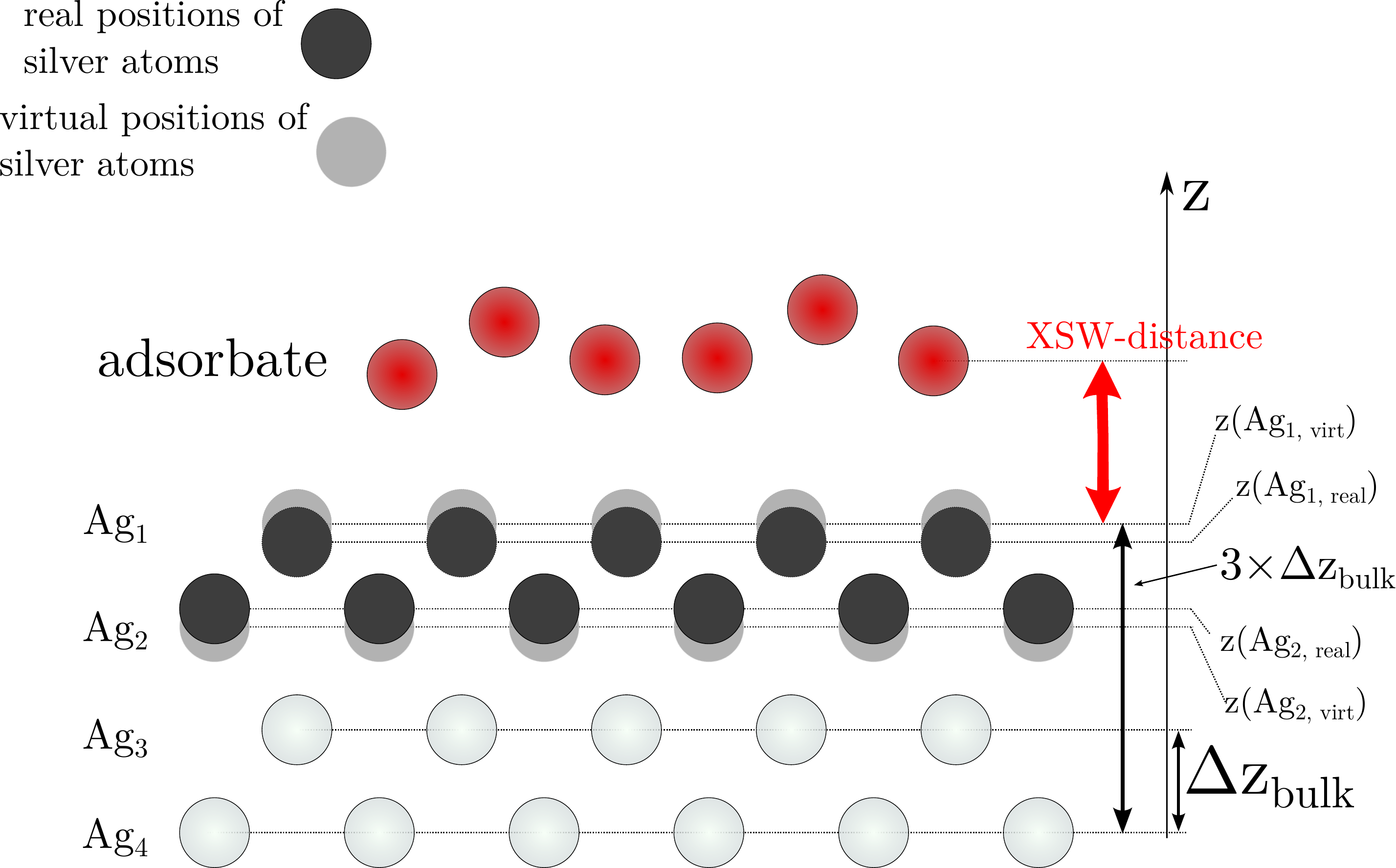}
\caption{XSW distances for four layer Ag(110) clusters. The relaxed (bulk)
  first and second layers are denoted by Ag$_{1,real}$ (Ag$_{1,virt}$)
  and Ag$_{2,real}$ (Ag$_{2,virt}$), respectively. $\Delta z_{bulk}$ is the bulk
  interlayer separation. }
\label{XSWD-definition}
\end{figure}

The distance of the atoms of an adsorbate can be determined experimentally
with the XSW technique which provides structural information of all those
atoms that give rise to a distinguishable signal in the X-ray photoelectron
spectrum (XPS). XSW determines the distance between these atoms and the
virtual plane of the relaxation-free uppermost layer of crystal atoms, i.e.,
the plane of the surface atoms if these did not relax from their bulk
positions \cite{Woodruff_2005_XSW_Review,Ton16}. 
According to the proposal of Woodruff
\cite{Woodruff_2005_XSW_Review} (see Fig.~\ref{XSWD-definition}) 
we calculate the XSW distance
of the atom $A$ in the molecule from the slab or cluster structure
as
\begin{align}
\label{eq:Def_XSW_big}
d_{A}&=(z_{\text{A}}-z_{\text{Ag}_n})-(n-1)\Delta z_{bulk},
\end{align}
where $z_{A}$ represents the vertical distance of the atom $A$ from the
surface, $n$ is the number of Ag-layers in the cluster, and $\Delta z_{bulk}$
is the distance obtained from periodic slab supercell calculations.
In eq.~\eqref{eq:Def_XSW_big} we rather use the theoretically
optimised than the experimental value for $\Delta z_{bulk}$ as the property of
interest is actually the position of the molecule on the top layer. The
position of this top layer is incorrectly described with the experimental
$\Delta z_{bulk}$ and this error increases with the number of layers in the slab
or cluster model.

\section{Results and discussion}

\subsection{Bulk Distances}
\label{sec:bulk-distances}


\begin{table*}
  \centering
  \caption{Next neighbor Ag-Ag distances, $r_{nm}$, between  Ag atoms in the
    $n$th and  the $m$th layer for the free Ag(110) surface and vertical
    distances of the Ag layers  as obtained with the slab calculations. 
    Ag$_{n}$ denotes an atom in the $n$th layer where $n=1$ is the topmost
    layer at the surface. All distances in \AA. Numbers in brackets are the
    percentage relaxations of the vertical layer distances.} 
  \label{tab:silver-distances}
  \begin{tabular}{lcccrrrrr}
\toprule
method & $r_{Ag_{1}-Ag_{2}}$&$r_{Ag_{1}-Ag_{2}}$&$ r_{Ag_{bulk}-Ag_{bulk}}$&
\multicolumn{2}{c}{$z_{Ag_1}-z_{Ag_2}$}&
\multicolumn{2}{c}{$z_{Ag_2}-z_{Ag_3}$}&
$\Delta z_{bulk}$\\
\colrule
PBE        & 2.864 &2.969 &2.933 &1.323&($-$9.8 \%) &1.537&($+$4.8 \%) &1.467\\
PBE-D2     & 2.847 &3.026 &2.921 &1.306&($-$10.6 \%) &1.661&($+$13.7 \%) &1.461\\
PBE-D3(BJ) & 2.836 &2.928 &2.880 &1.350&($-$6.3 \%) &1.534&($+6.5 \%$) &1.440\\
expt.\footnote{Ref. \cite{ag-gitterkonstante}}&&&2.889&&&1.445\\
\botrule
  \end{tabular}
\end{table*}

Tab.~\ref{tab:silver-distances} shows the nearest neighbour and adjacent
atomic layer distances of silver atoms as obtained in the slab calculations
with the noD, D2, and D3(BJ) dispersion corrections as well as the
experimental bulk distances. The nearest neighbour distance obtained with the
D3(BJ) approach deviates only by \SI{-0.009}{\angstrom} from the experimental
value while the two other approaches overestimate the Ag-Ag distances by about
\SI{0.03}{\angstrom}.

All methods predict that the distance between the second and third layers is
longer than the layer distance in the bulk, $\Delta z_{bulk}$, while the
distance between the first and second layer atoms is smaller than that.
LEED-$I(V)$ experiments of Nascimento {\it et al.}~\cite{Nas03} as well as
XSW data of Bauer {\em et al.}~\citep{PhysRevB.86.235431} and DFT
calculations~\cite{Nas03,Wan01a,Nar02} indicate that the distance between the
first and second atomic layer of silver is contracted by $-7.5\pm 3.0$~\% while
$d_{{\rm Ag_{2}-Ag_{3}}}$ is extended by $2\pm5$~\%  with respect to the bulk
interlayer distance $\Delta z_{bulk}=\SI{1.445}{\angstrom}$.

\begin{figure}
\centering
\includegraphics[width=\columnwidth]{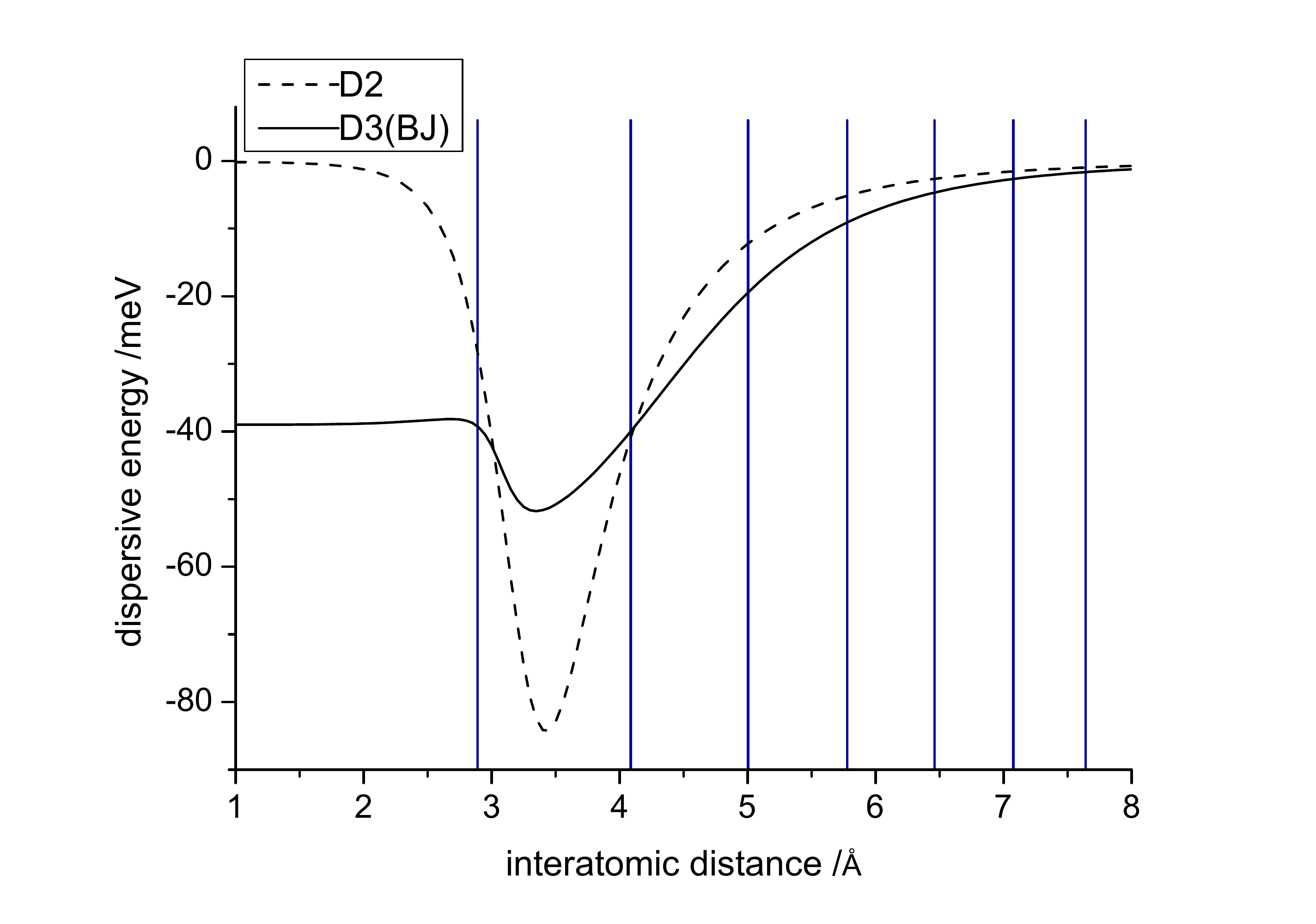}
\caption{D2- and D3(BJ)-Dispersion contributions between two silver atoms as a
  function of the inter atomic distance. The blue vertical lines indicate
  (from left to right) the experimental next neighbour, second next neighbour,
  etc.\  distances in the bulk silver structure.
  \label{fig:dispersion-contribution-d2-d3bj} }
\end{figure}

We note that this relaxation pattern is not observed in the D2 and D3(BJ)
model. This is probably due to the repulsive part of the interaction energy
that appears from the damping of the dispersion correction.
Fig.~\ref{fig:dispersion-contribution-d2-d3bj} shows that this damping leads
to artificial repulsive contributions to the total energy for the typical
nearest neighbour silver-silver distance which are, thus, stretched if these
dispersion corrections are applied.

\subsection{Adsorbate Structures}
\label{sec:adsorbate-structures}

\begin{figure}[htbp]
\includegraphics[width=0.7\columnwidth]{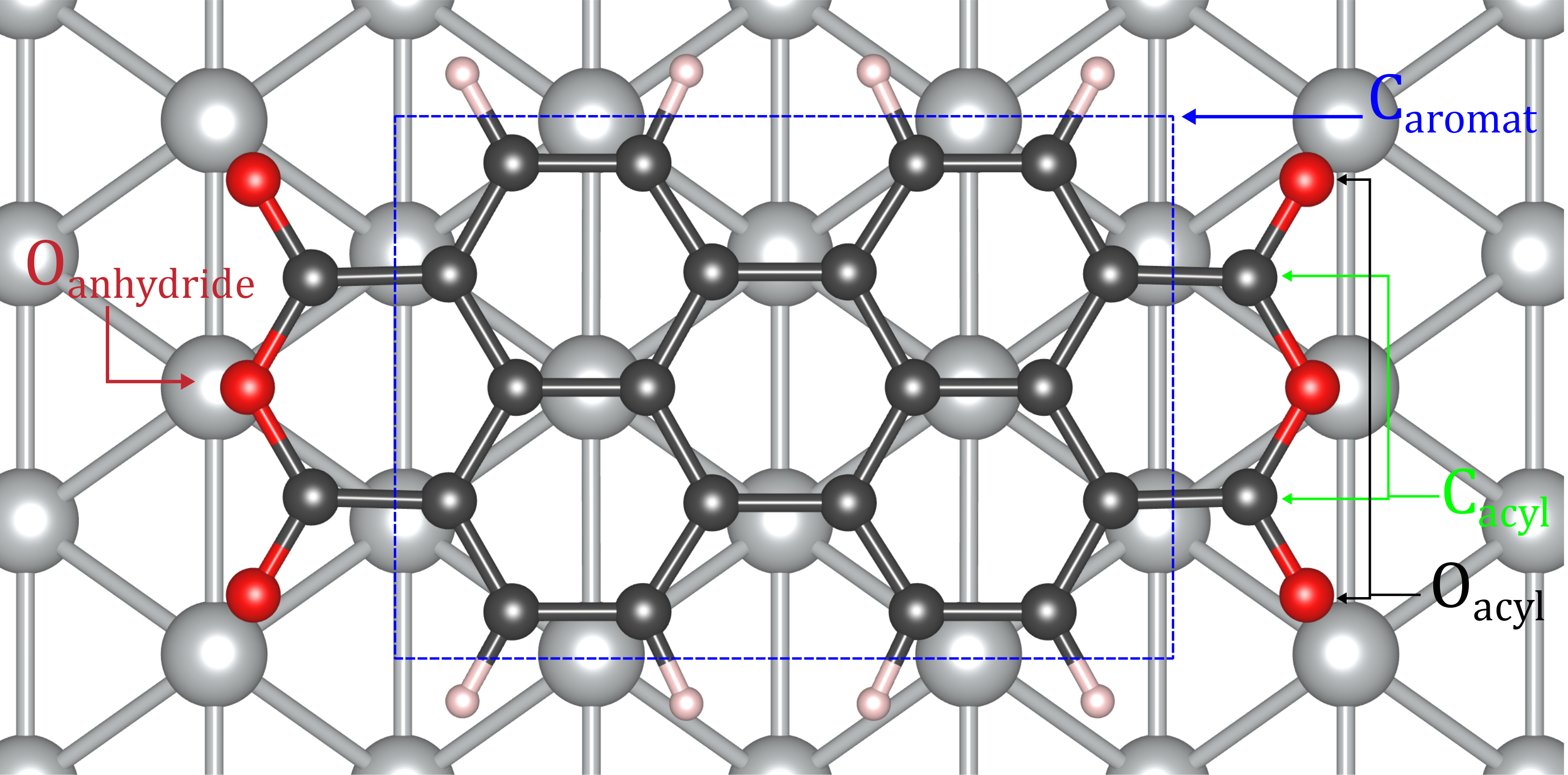}
\caption{PTCDA molecule on the Ag(110) surface; the perylene core is
  represented as the C$_\text{aromat}$, the acyl carbon atoms are marked as
  C$_\text{acyl}$, the anhydride oxygen as O$_\text{anhydride}$ and the acyl
  oxygen as O$_\text{acyl}$. \label{fig:atom-definitions-and-convergence}}
\end{figure}
In this section the convergence of the adsorbate structures is investigated
with respect to (i) the dispersion scheme [noD, D2, D3(BJ)], (ii) the basis
set [def-SV(P), def2-TZVP], (iii) fixing of border atoms [fix2, fix12], and
(iv) the cluster size (32, 50, 82, 170, 218, 226 and 290 silver atoms). Most
of these  combinations were investigated and compared with the results
of the slab-supercell calculations. As the Ag$_{170}$ cluster with the
fix12-scheme and the def2-TZVP basis set provides reasonably well converged
results, we shall use this as a reference for the following comparisons.
Further results are collected in the supporting information of this article.

The adsorption height of PTCDA above the Ag(110) surface is determined by the
following XSW distances: C$_\text{aromat}$ is the {\textit{average}} distance of
the 20 carbon atoms in the perylene core (see Fig.~\ref{figure1}) while
C$_\text{acyl}$, O$_\text{anhydride}$, and O$_\text{acyl}$ denote the distances of
the respective carbon and oxygen atoms (see Figs.~\ref{XSWD-definition} and
\ref{fig:atom-definitions-and-convergence}).


\begin{table}
  \centering
  \caption{XSW distances of the Ag170 cluster for different  dispersion
    corrections as obtained with the fix12 scheme and the
    def2-TZVP basis. Distances are given in \r A.} 
  \begin{tabular}{cdddd}
    \toprule
    dispersion scheme & 
\multicolumn{1}{l}{O$_\text{acyl}$}  & 
\multicolumn{1}{l}{O$_\text{anhydride}$} & 
\multicolumn{1}{l}{C$_\text{acyl}$}  & 
\multicolumn{1}{l}{C$_\text{aromat}$}   \\ 
		\colrule
noD    & 2.252 & 2.351 & 2.444 & 2.717\\ 
D2     & 2.277 & 2.370 & 2.434 & 2.666\\ 
D3(BJ) & 2.311 & 2.411 & 2.468 & 2.638\\ 
\colrule
slab D3(BJ)\footnotemark[1]  & 2.353   & 2.454 & 2.509   & 2.642  \\
slab D3(BJ)\footnotemark[2]  & 2.38   & 2.46   & 2.53    & 2.71 \\
expt.\footnotemark[3] & 2.32& 2.41& 2.45& 2.59 \\
\botrule
\end{tabular}
\label{tab:TableDisp}
\footnotetext[1]{this work}
\footnotetext[2]{{Ref. \cite{Mer13a}}}
\footnotetext[3]{{Ref. \cite{PhysRevB.86.235431}}}
\end{table}
In Table~\ref{tab:TableDisp} we present the effect of changing the dispersion
correction scheme. XSW distances of the Ag170 cluster are compared with the
slab results of this work and of Ref.~\cite{Mer13a} as well as with the
experimental results of Bauer {\em et al.}~\cite{PhysRevB.86.235431}. As also
pointed out in prior work
\cite{Scholz_short_paper,Scholz_2009_long_paper,PhysRevLett.99.176401,Tkatchenko_2010}
a proper inclusion of the dispersion interaction is of crucial importance for
predicting a reliable structure for organic molecules upon a metal surface.
Neglect of the dispersion interaction leads to much too short XSW distances of
the oxygen atoms (about \SI{0.07}{\angstrom} shorter than the experimental
values) while for the aromatic carbon atoms the corresponding values of the
cluster are \SI{0.11}{\angstrom} larger than the experimental ones. Thus, if
dispersion interaction is not taken into account, the organic molecule is
predicted to be substantially buckled upon the surface due to covalent and
ionic bonds between the metal and the electronegative oxygen atoms
\cite{PhysRevB.86.235431}. This buckling is diminished by dispersion
interactions which provide a bonding mechanism between the carbon and silver
atoms. Thus, while dispersion is not necessarily improving the description of
the metal surface structure, it is mandatory for a realistic representation of
the organic substrate upon the metal surface.

As already seen for the metal structure, Tab.~\ref{tab:TableDisp} shows that
the structure of PTCDA on the metal is better predicted by the more advanced
D3(BJ) scheme than by its predecessor D2. The latter tends to underestimate
the XSW distances of the oxygen atoms by \SI{0.04}{\angstrom} while the
corresponding values for the aromatic carbon atoms are predicted to be
\SI{0.07}{\angstrom} larger than the experimental values. In general the D2
dispersion correction performs better than the pure PBE approach but the
D3(BJ) results are significantly better. Thus, we shall only discuss the
latter in the following.


\begin{table}
  \centering
  \caption{XSW distances in \AA\ upon variation of border fixing schemes for
    two different cluster models obtained with the def2-TZVP basis and the
    slab results. The D3(BJ) dispersion correction was used in all
    calculations. \label{tab:TableFix} }  
  \begin{tabular}{ccdddd}
    \toprule
    cluster &fix status & 
\multicolumn{1}{l}{O$_\text{acyl}$}  & 
\multicolumn{1}{l}{O$_\text{anhydride}$} & 
\multicolumn{1}{l}{C$_\text{acyl}$}  & 
\multicolumn{1}{l}{C$_\text{aromat}$}   \\ 
\hline
    170 & fix2  & 2.326 & 2.429 & 2.483 & 2.644\\
        & fix12 & 2.311 & 2.411 & 2.468 & 2.638\\
    290 & fix2  & 2.335 & 2.438 & 2.485 & 2.623\\
        & fix12 & 2.334 & 2.441 & 2.483 & 2.616\\
\botrule
\end{tabular}
\footnotetext[1]{this work}
\footnotetext[2]{{Ref. \cite{Mer13a}}}
\footnotetext[3]{{Ref. \cite{PhysRevB.86.235431}}}
\end{table}
Keeping all other parameters fixed and varying the border fixing scheme for
the 170 atom cluster (Ag170) we see that the {\fixmacro{fix2}} (fixing the
border atoms only in the second layer) scheme provides slightly larger XSW
values than the {\fixmacro{fix12}} (fixing the border atoms of the first and
second layers) scheme (Table~\ref{tab:TableFix}). For the Ag170 cluster a
change from the fix2 to the fix12 scheme causes an increase of all XSW values
by 0.007--\SI{0.018}{\angstrom} while for the biggest cluster, Ag290, the
different schemes of fixing the border atoms has an almost negligible effect
($<\SI{0.003}{\angstrom}$) on the XSW distances (Table~\ref{tab:TableFix}). In
the following, we choose the {\fixmacro{fix12}} scheme
as the atoms far away from the molecule will not show significant displacement
from their bare surface like positions.

\begin{table}
  \centering
  \caption{XSW distances obtained for the Ag170 cluster and various basis sets. 
    The fix12 scheme and the D3(BJ)  dispersion correction were used for the
    cluster calculations. Distances are given in \AA.\label{tab:TableBasis}}   
  \begin{tabular}{cdddd} 
    \toprule
    basis set & 
\multicolumn{1}{l}{O$_\text{acyl}$}  & 
\multicolumn{1}{l}{O$_\text{anhydride}$} & 
\multicolumn{1}{l}{C$_\text{acyl}$}  & 
\multicolumn{1}{l}{C$_\text{aromat}$}   \\ 
    \colrule
    def-SV(P)                & 2.378 & 2.451 & 2.537 & 2.728 \\
    def2-TZVP                & 2.311 & 2.411 & 2.468 & 2.638 \\
    partly def2-QZVP\footnote{def2-QZVP basis for PTCDA
      and first Ag layer atoms. def2-TZVP for other atoms.}
                             & 2.292 & 2.391 & 2.453 & 2.631 \\
    \colrule
    slab\footnote{this work} & 2.353 & 2.454 & 2.509 & 2.642 \\
    \botrule
  \end{tabular}
\end{table}
In Table~\ref{tab:TableBasis} the XSW results for different basis sets and the
Ag170 cluster are shown. With the def-SV(P) basis, the XSW distances of the
acyl carbon and oxygen and for the anhydride oxygen -- the functional atoms --
are in good agreement (within \SI{0.03}{\angstrom}) with the slab-supercell
approach but the deviation of the aromatic carbon is appreciably larger (0.09
\AA). A much more balanced structure is obtained with the def2-TZVP basis
which underestimates the XSW distances of the functional atoms by
0.03--\SI{0.04}{\angstrom} but exactly reproduces the positions of the
aromatic carbon atoms. This shows that the def2-TZVP is, as expected, a
preferred choice over the def-SV(P) basis. With the larger mixed basis (partly
def2-QZVP) all XSW distances decrease by 0.01--\SI{0.02}{\angstrom}. All our
results presented henceforth are performed with the def2-TZVP basis, which
represents a good compromise between accuracy and efficiency.

\begin{table}
  \centering
  \caption{XSW distances obtained for different cluster sizes, slab models and
    experimental data. 
    The fix12 scheme, the def2-TZVP basis  and the D3(BJ)  dispersion
    correction was used for the 
    cluster calculations. Distances are given in \AA, adsorption energies,
    $E_\text{ads}$, in eV. The numbers in brackets 
    are published  error estimates. } 
  \label{tab:geoconvergence}
  \begin{tabular}{ldddde}
\toprule
    cluster size & 
\multicolumn{1}{l}{O$_\text{acyl}$}  & 
\multicolumn{1}{l}{O$_\text{anhydride}$} & 
\multicolumn{1}{l}{C$_\text{acyl}$}  & 
\multicolumn{1}{l}{C$_\text{aromat}$}    & 
\multicolumn{1}{l}{$E_\text{ads}$}\\
\colrule
32  & 2.366  & 2.568  & 2.542  & 2.639 & -4.519 \\
50  & 2.421  & 2.579  & 2.561  & 2.620 & -4.312 \\
82  & 2.423  & 2.585  & 2.563  & 2.604 & -4.176 \\
170 & 2.311  & 2.411  & 2.468  & 2.638 & -4.350 \\
218 & 2.314  & 2.430  & 2.469  & 2.610 & -4.433 \\
226 & 2.358  & 2.448  & 2.497  & 2.631 & -4.326 \\
290 & 2.334  & 2.441  & 2.483  & 2.616 & -4.504 \\
\colrule
slab\footnotemark[1] & 2.353 & 2.454 & 2.509 & 2.642 & -4.90 \\
slab\footnotemark[2] & 2.38 & 2.46 & 2.53 & 2.71 & 
\multicolumn{1}{c}{$-$4.4\footnotemark[3]}\\
expt.\footnotemark[2] & 
\multicolumn{1}{c}{2.30(4)}& 
\multicolumn{1}{c}{2.38(3)}&
\multicolumn{1}{c}{ 2.45(11)}&
\multicolumn{1}{c}{ 2.58(1)} & 
\multicolumn{1}{c}{} \\
expt.\footnotemark[4] & 
\multicolumn{1}{c}{2.32(5)}& 
\multicolumn{1}{c}{2.41(6)}& 
\multicolumn{1}{c}{2.45(11)}& 
\multicolumn{1}{c}{2.59(1)}& 
\multicolumn{1}{c}{} \\
\botrule
  \end{tabular}
\footnotetext[1]{this work}
\footnotetext[2]{Ref.~\cite{PhysRevB.86.235431}}
\footnotetext[3]{estimated from Fig. 4 in Ref.~\cite{PhysRevB.86.235431}}
\footnotetext[4]{Ref. \cite{Mer13a}}
\end{table}
In Tab.~\ref{tab:geoconvergence}, the XSW distances obtained for different
cluster sizes with the def2-TZVP basis, the fix12 border atom fixing and the
D3(BJ) dispersion correction are collected together with periodic slab and
experimental results. The trend of the XSW distances of the respective atoms
O$_\text{acyl}<$O$_\text{anhydride}$$<$C$_\text{acyl}$$<$C$_\text{aromat}$ is
obtained from the periodic calculations and for all clusters shown in
Tab.~\ref{tab:geoconvergence}. However, this trend is not reproduced in all
cases. In particular, the D2 dispersion scheme provides erratic substrate
structures if small cluster sizes are {\employ}ed (see supporting information
for further details).

We see that the XSW distances essentially converge to the periodic results as
the size of the clusters increase. The XSW distances of the Ag170 cluster are
in significantly better agreement with the slab-supercell results if compared
with the Ag32, Ag50 and Ag82 clusters. However, the XSW distances of this
cluster are by about \SI{0.03}{\angstrom} shorter than the corresponding slab
values. Increasing the cluster size tends to increase the XSW distances to the
oxygen and the C$_\text{acly}$ atoms while the C$_\text{aromat}$ distances
decrease. We note that the change of these distances is not a smooth function
of the number of atoms in the cluster as the Ag226 cluster predicts XSW
distances that are about by \SI{0.03}{\angstrom} larger than the corresponding
values of the Ag170, Ag218 and Ag290 clusters.

For the larger clusters (consisting of 170, 218, 226 and 290 Ag atoms) with
D3(BJ) dispersion correction and def2-TZVP basis, the cluster results deviate
only up to \SI{0.03}{\angstrom} from the periodic slab-supercell approach of
the present work while the theoretical results of Bauer {\em et
  al.}~\cite{PhysRevB.86.235431} provide XSW distances which are
0.03--\SI{0.09}{\angstrom} larger. We deduce that a metallic cluster should be
able to retain properties of a true adsorbate molecule upon a metal surface if
the metal cluster contains at least one row of metal atoms in lateral
extension beyond the atoms that are in direct contact to the adsorbate
molecule. However, the cluster approach itself represents an error source that
limits the accuracy of the determined structure. For the larger cluster models
the magnitude of these variations is smaller than uncertainties due to the
electronic structure method, the dispersion correction, or the basis set.


The computed adsorption energies are also displayed in
Tab.~\ref{tab:geoconvergence}. The convergence of the cluster adsorption
energies with the cluster size is comparable to the convergence of the XSW
distances. A value of $-4.4\pm$\SI{0.1}{eV} can be deduced from the variation
of the adsorption energy for the clusters with 170 and more silver atoms.
While this is in perfect agreement with the periodic slab results of Bauer
{\em et al.}~\cite{PhysRevB.86.235431} our own slab calculations provide a
lower adsorption energy of \SI{-4.90}{eV}. Since our computational parameters
are very similar to Ref.\cite{PhysRevB.86.235431}, this discrepancy requires
further investigation.

\section{Conclusions and outlook}

We have investigated the applicability of the cluster approach for predicting
the adsorbate structure of an organic molecule upon metallic surfaces. PTCDA
on the Ag(110) surface was chosen for this comparison as it is experimentally
well investigated and the standard theoretical approach (periodic slab
supercell calculations) has been applied to this system. We chose
to describe the electronic structure at the DFT level with the PBE functional.

Our results demonstrate that a reliable description of the surface substrate
interaction requires several points to be addressed
\begin{itemize}
\item Dispersion interactions have to be added to the DFT approach preferably
  with the D3(BJ)~\cite{Grimme_2011_damping_function} scheme which was found
  to be significantly better than the older D2~\cite{Grimme_D2_2006} variant.
  The latter tends to give rise to artificial surface reconstruction due to
  the damping of the dispersion interaction for short atomic distances which
  are ameliorated in the D3(BJ) approach.
\item For the cluster approach sufficiently large basis sets are needed to
  converge the adsorbate properties. We found that the Ahlrichs triple-$\zeta$
  quality basis set def2-TZVP is necessary for structural convergence within
  \SI{0.02}{\angstrom}.
\item A consistent setup of the metal cluster turns out to be a nontrivial
  task. 
We deduce that a metallic cluster should be able to retain properties of a
true adsorbate molecule upon a metal surface if the metal cluster contains at
least one row of metal atoms in the lateral
%
  extension beyond the atoms that are in direct contact
  to the adsorbate molecule and a sufficient number of metal atom layers.
  Furthermore, during structure optimisations it is important to fix the
  positions of atoms at the border of the cluster to their positions at the
  metal surface.
\item Several ``tricks'' are required to make the cluster calculations
  feasible. Among them are the RI~\cite{Vah93,Eichkorn_1997} and
  MARI-J~\cite{Sierka_2003} approximations to accelerate the computations, a
  Fermi smearing to simplify the determination of occupations of orbitals near
  to the Fermi energy and appropriate damping schemes for the slowly
  converging SCF iterations.
\end{itemize}

The cluster approach provides a converged adsorption energy of PTCDA on
Ag(110) of $-4.4\pm$\SI{0.1}{eV} for the given DFT and dispersion method and
the XSW distances of PTCDA on Ag(110) were determined with an accuracy of
about \SI{0.03}{\angstrom}. The XSW distances of the larger clusters agree
with those of the periodic slab calculations within this error limit. The
variation of results from the cluster model indicate a limitation of this
approach even for rather large cluster sizes. However, the ``cluster-error''
for XSW distances is in the same order of magnitude as 
experimental errors ($\approx$\SI{0.05}{\angstrom}~\cite{Ger14})
and
smaller than the mean absolute error of \SI{0.06}{\angstrom} reported for the
difference between theoretical and experimental XSW distances in the very
recent work of Maurer {\em et al.}~\cite{Mau16}.

The present work indicates that it may be possible to design cluster methods
that predict structural properties of organic adsorbates upon metal surfaces
with a similar accuracy as experimental results. Such a protocol would be very
useful as a computationally comparable or even favourable alternative to the
well established periodic slab approach. This would have particular advantages
for cases where periodic symmetry is absent due to the structure of the
investigated system or for other cases where the periodic slab approach is not
applicable. Last but not least, a validated finite cluster ansatz would
provide an independent theoretical access to organic adsorbates on metal
surfaces. 

\section{Acknowledgements}

We acknowledge support by the Volkswagen-Stiftung (BE and RFF) and the
Deutsche Forschungsgemeinschaft DFG in the framework of the Research Training
Group 1221 ``Control of the Electronic Properties of Aggregated
$\pi$-conjugated Molecules'' (VS and BE) as well as the research project
Fi620/3-1 (RFF and HGM) and the SFB 1083 ``Structure and Dynamics of Internal
Interfaces'' (RT). We are grateful for compute resources of the Centre
for Light-Matter Interaction, Sensors \& Analytics (LISA+) of the University
of T\"ubingen and the HLR Stuttgart.

\bibliography{Literaturejb}

\end{document}